
\overfullrule0pt

 \newtoks\slashfraction
 \slashfraction={.13}
 \def\slash#1{\setbox0\hbox{$ #1 $}
 \setbox0\hbox to \the\slashfraction\wd0{\hss \box0}/\box0 }

  \def\Buildrel#1\under#2{\mathrel{\mathop{#2}\limits_{#1}}}

\def\lozenge{\boxit{\hbox to 1.5pt{%
             \vrule height 1pt width 0pt \hfill}}}

\def\H{{\cal H}}
\def\K{{\cal K}}
\def\L{{\cal L}}
\def\M{{\cal M}}

\def\P{{\cal P}}

\def\mn{{\mu\nu}}

 \doublespace
 \pubnum{6392}
\date{September 1993}
 \pubtype{T/E}
 \titlepage
 \title{Quantization of Gauge Field Theories on the Front-Form
 without Gauge Constraints I : The Abelian Case
 \doeack}
 \author{ Ovid C. Jacob\foot{jacob@slacvm.slac.stanford.edu}
 }
 \SLAC

\abstract
Recently, we have
 proposed   a new front-form quantization
which treated both the $x^{+}$ and the $x^{-}$  coordinates as
          front-form   'times.' This
quantization  was found to preserve parity explicitly.
In this paper we extend this construction to local Abelian
gauge fields .    We quantize        this theory using
a method     proposed originally by Faddeev and Jackiw   .
We emphasize here the feature that quantizing along both $x^+$ and
$x^-$ , gauge theories does not require extra constraints (also known as
'gauge conditions') to determine the solution uniquely.
\vfill
\centerline{Submitted to Phys. Rev. D}
\chapter{Introduction}
Front-form quantization is usually done by quantization along
the front $x^+=const  $ .
Usually this is done by quantizing a system with constraints
\REF\cd{Kurt Sundermeyer, 'Constrained Dynamics', Springer-Verlag,
New York, 1982 }
\refend
\REF\gt{D. M. Gitman and I. V. Tyutin, 'Quantization of Fields with
Constraints', Springer-Verlag, New York, 1990}
\refend
\REF\gov{Jan Govaerts , 'Hamiltonian Quantization and Constrained
Dynamics', Leuven University Press, Leuven, Belgium, 1991}
\refend
\REF\ht{Marc Henneaux and Claudio Teitelboim, 'Quantization of Gauge
Systems', Princeton University Press, Princeton, NJ, 1992}
\refend .
In a previous papers
\REF\pclcq{Ovid C. Jacob, 'Parity Conserving Light-Cone Quantization',
SLAC-PUB-6188,May 1993, hep-th/9305076, Submitted to Mod. Phys. Lett A.}
\refend
\REF\pffq{Ovid C. Jacob
'Parity and Front-Form quantization of Field Theories,' SLAC-PUB
August 1993 }
\refend
\REF\dm{see also Dharam V. Ahluwalia and Mikolaj Sawicki,
'Spinors on the Front-Form,' LANL preprint, October 1993 }
\refend
we   introduced a quantization which treated $x^+$ and
$x^-$  on equal footing.
\REF\other{Couple of recent preprints also address the issue of
consistency of field equations on light-cone (T. Heinzl and E.
Werner, Regensburg preprint TPR-93-3) or for null-plane field
theory (Norbert E. Ligterink and B.L.G. Bakker, Vrije Universiteit,
Amsterdam preprint, 1993) . In both of these papers, the authors
stay in the usual approach of taking only {\bf one} light-cone
(null-plane) 'time.' The Regensburg paper actually makes some rather
strong claims (with scant support)
regarding lack of need for a second light-like
hyperplane, but they do it in the context of a mixed initial-boundary
value problem, whereas here we consider an initial value problem.}
\refend
The main argument given was that
this new approach was  manifestly parity invariant.
We also pointed out that this new approach had the same number
of degrees of freedom as the equal-time approach.

We'd like to point out that in some work involving initial value problems
in gravity using front-form coordinates
\REF\ivpa{R. Gambini, A. Restuccia, Phys. Rev. \us{D17},
3150, (1978)}
\refend
\REF\ivpb{R. K. Sachs, J. Math. Phys. \us{3}, 908, (1962)}
\refend
\REF\ivpc{R. Penrose, Gen. Relat. and Grav. \us{12}, 225, (1980);
Roger Penrose and Wolfgang Rindler, 'Spinors and Space-Time,' vol. I,
Chapter 5 ,Cambridge University Press, New York, 1984 }
\refend
\REF\ivpd{H. M\"uller zum Hagen and H-J Seifert,
Gen. Relat. and Grav. \us{ 8}, 259, (1977)}
\refend
\REF\ivpe{H. Bondi, M. G. J. van der Burg and A. W. K. Metzner,
Proc. R. Soc. London, \us{A269}, 21, (1962)}
\refend  ,
the initial data for these coordinates is
also specified along {\bf both} $x^+ = const$ and $x^- = const$
surfaces as well as at $x^+ =x^- = 0$ .
  R. Penrose [\ivpc]
points out that in this approach there are {\it no} constraints  .
In this paper we study how this approach could
bypass much of the difficulty coming from the presence of constraints
in the presence of local Abelian ($U(1)$)  gauge symmetry.
(A future paper will look at the non-Abelian   $(SU(N) )$ case.
The point is as follows: in usual gauge theory quantization, the
gauge condition is a relation (constraint) between the quantizing
degrees of freedom (the initial data). We show in this work that
using the two null hyperplanes, we don't need any constraints between
initial data.

There are two points which we should mention. First, the use of
the reduced phase space quantization of Faddeev and Jackiw
\REF\fj{L. D. Faddeev and R. Jackiw,  Phys. Rev. Lett. \us{60},
1692,(1988)}
\refend
(see also
\REF\jack{ R. Jackiw, '(Constrained) Quantization without Tears,'
CTP-\#2215, hep-th/9306075}
\refend  )
 allows us to get the commutation relations easily. Second, as they
point out, if the two-form  (which goes in defining the
equations of motion ) is invertible, then there are {\bf no
constraints} . This fact  coupled with Penrose's remark [\ivpc]
seem to imply that using two null hyperplanes, we always have
an invertible two-form , hence never any constraints. Obviously
this greatly facilitates the quantization procedure.
\chapter{Reduced Phase Space Quantization of QED   }
We follow the reduced phase space quantization of Faddeev and
Jackiw [\fj] to study $QED$ :
$$\L =-{1\over2}F_{\mn} F^{\mn}
+ {i\over 2} \partial^{\nu}\overline{\psi} \gamma_{\nu} \psi
- \overline{\psi} {i\over 2} \partial^{\nu}\gamma_{\nu} \psi
 -m \overline{\psi} \psi  + \L_I \eqn\f $$
where $\L_I $ is the interaction part
of the Lagrangean:
$$ \L_I = -e  \overline{\psi}\gamma^\nu \psi A_\nu \eqn\li $$
The Euler-Lagrange equations of motion are
$$ \partial_\mu F^\mn - e \overline{\psi}\gamma^\nu \psi =0\eqn\ema$$
$$(i\gamma_\mu\partial^\mu -m -e\gamma_\mu A^\mu)\psi = 0\eqn\empsi $$

To obtain the Hamiltonians for evolution along $x^+$ and $x^-$, we
follow the approach of previous papers [\pclcq] and [\pffq] , so
we write  $\L$ out explicitly :
$$\L d^4 x = \big\{-{1\over 2} F_{ij} F^{ij} - F_{+-}F^{+-}-
F_{+i}F^{+i}- F_{-i}F^{-i}   $$
$$ -2e \psi^{\dagger}_+ A_+ \psi_+ -
2e \psi^{\dagger}_- A_- \psi_- +
e\psi^{\dagger}_+\gamma_0\gamma_i A^i \psi_- +
e\psi^{\dagger}_-\gamma_0\gamma_i A^i \psi_+ +
\psi^{\dagger}_+ {i\partial^- \over 2}\psi_+ - {i\over 2}
(\partial^- \psi^{\dagger}_+ )\psi_+   $$
$$ +\psi^{\dagger}_- {i\partial^+ \over 2}\psi_- - {i\over 2}
(\partial^+ \psi^{\dagger}_- )\psi_- +
-\psi^{\dagger}_+\gamma_0\gamma_i {i\partial_i \over 2}\psi_- +
{i\over 2}(\partial_i\psi^\dagger_-)\gamma_0\gamma_i\psi_+  $$
$$ -\psi^{\dagger}_-\gamma_0\gamma_i {i\partial_i \over 2}\psi_+ +
{i\over 2}(\partial_i\psi^\dagger_+)\gamma_0\gamma_i\psi_- -
m\psi^{\dagger}_+{\gamma^- \over 2}\psi_- -
m\psi^{\dagger}_-{\gamma^+ \over 2}\psi_+ \big\}
{dx^- dx^+ d^2 x_\perp \over 2}    \eqn\fg$$
where $\psi_\pm = \Lambda_\pm \psi$ and $\Lambda_\pm={1\over 2}\gamma^0
\gamma^\pm$ . Note also that
$\partial_\nu = {\partial \over \partial x^\nu}$ so that
$\partial^- = 2 \partial_ + = 2 {\partial \over \partial x^+ }$ and
$\partial^+ = 2 \partial_ - = 2 {\partial \over \partial x^- }$ .
The corresponding  conjugate momenta for     $x^+$-derivatives are
$$ \pi_A^i   =  {   \partial \L \over \partial (\partial^{-} A_i    ) } =
-{1\over 2}F^{+ i }  \eqn\pa $$
$$\pi_{\psi}(x)= {\partial \L \over \partial (\partial^{-} \psi) } =
 {i\over 2}\psi^{\dagger}_{+} \eqn\i$$
$$\pi_{\psi^{\dagger}}(x)={\partial \L \over
  \partial (\partial^{-} \psi^{\dagger}) }=-{i\over 2}\psi_{+} \eqn\j$$
For the momenta corresponding to $x^-$-derivatives we get similar forms:
$$\rho_A^i  =  {   \partial \L \over \partial (\partial^{+} A_i    ) } =
- {1\over 2} F^{- i  }  \eqn\hh$$
$$\rho_{\psi}(x)={\partial \L \over  \partial (\partial^{+} \psi) } =
 {i\over 2}\psi^{\dagger}_{-} \eqn\k$$
$$\rho_{\psi^{\dagger}}(x)=
{\partial \L \over \partial (\partial^{+} \psi^{\dagger}) }=
-{i\over2}\psi_{-} \eqn\l$$
We rewrite $\L d^4 x$ in the following way
$$\L d^4 x = {1\over 2}\big\{\pi_A ^i \partial^- A_i +
\pi_A^i \partial^- A_i + \rho_A ^i \partial^+ A_i +
\rho_A^i \partial^+ A_i  $$
$$+\pi_\psi\partial^-\psi_+  +\pi_\psi\partial^-\psi_+  +
(\partial^-\psi^{\dagger}_+)\pi_{\psi^{\dagger}}
   +(\partial^-\psi^{\dagger}_+)\pi_{\psi^{\dagger}} $$
$$ +\rho_\phi\partial^+\phi + \rho_\phi\partial^+\phi +
    \rho_\psi\partial^+\psi_- +\rho_\psi\partial^+\psi_-  +
(\partial^-\psi^{\dagger}_-) \rho_{\psi^{\dagger}} +
(\partial^-\psi^{\dagger}_-) \rho_{\psi^{\dagger}} \big\}d^4 x $$
$$ -\H dx^+  -\K dx^- +\M d^4 x \eqn\fjj$$
The meaning of these terms is as follows : the first bracket represents
the p-q-dot terms which go into the definitions of the canonical
commutation relations; the second and third term are the Hamiltonians
which  define the evolution of the system along $x^+$ , given by $\H$,
and along $x^-$ given by  $\K$
  finally, the last term contains
the remaining
pieces which give the 'constraints', though are not  'true constraints'
[\fj] as are consistent with the gauge field
equations of motions. The
Hamiltonians $\H$ and $\K$ are :
$$\H = {dx^- dx^2_\perp \over 2}\big\{{1\over 2}(
B^{-2} )    +2e \psi^\dagger_+ A_+ \psi_+
            +2e \psi^\dagger_- A_- \psi_-
+\psi^{\dagger}_+\gamma_0\gamma_i {i\partial_i \over 2}\psi_-
+e\psi^{\dagger}_+\gamma_0\gamma_i  A^i \psi_- $$
$$ -{i\over 2}(\partial_i\psi^\dagger_-)\gamma_0\gamma_i\psi_
  -e\psi^\dagger_- \gamma_0\gamma_i A^i \psi_+
   +m\psi^{\dagger}_+{\gamma^- \over 4}\psi_- +
m\psi^{\dagger}_-{\gamma^+ \over 4}\psi_+\big \} \eqn\pmin$$
$$\K = {dx^+ dx^2_\perp \over 2}\big\{ {1\over 2}(
B^{+2} )    +2e \psi^\dagger_- A_- \psi_-
       +2e   \psi^\dagger_+ A_+ \psi_+
+\psi^{\dagger}_-\gamma_0\gamma_i {i\partial_i \over 2}\psi_
-e\psi^{\dagger}_-\gamma_0\gamma_i A ^i \psi_+  $$
$$ -{i\over 2}(\partial_i\psi^\dagger_+)\gamma_0\gamma_i\psi_-
  +e  \psi^\dagger_+ \gamma_0\gamma_i A^i \psi_-
   +m\psi^{\dagger}_-{\gamma^+ \over 4}\psi_+ +
m\psi^{\dagger}_+{\gamma^- \over 4}\psi_-\big \} \eqn\pplu$$
where $B^- = B^+ = {1\over\sqrt{2} }F^{12} $,
and for the 'constraints' we get whatever is left over

$$\M = \big\{ -\partial_i A_+ F^{+i} - \partial_i A_- F^{-i}-
F_{+-} F^{+-} +2 e A_+ \psi^\dagger_+ \psi_+ +2 e A_-\psi^\dagger_-
\psi_- \big\} \eqn\mt$$

Well, we can
rewrite is as ( up to total derivatives) :
$$\M = A_+ C_\pi + A_- C_\rho \eqn\amn$$
and the 'constraints' $C_\pi$ and $C_\rho$ are
$$C_\pi =-\partial_- F^{-+} -\partial_i F^{i+} +2 e \psi^\dagger_+
\psi_+ \eqn\cpi $$
$$C_\rho=-\partial_+ F^{+-} -\partial_i F^{i-} +2 e \psi^\dagger_-
\psi_- \eqn\crho$$
We see that $C_\pi = C_\rho = 0$ identically by the classical equation of
motion for the gauge fields , as in equation [\ema] for $\nu =+,-$.

Let us write the $\L dx^4$ with the explicit momenta dependence ( up
to total derivatives which we can discard [\fj],
\REF\zh{Wei-Min Zhang and Avaroth Harindranath ,'Light-Front QCD:
Role of Longitudinal Boundary Integrals', hepth@xxx/9302119;
Wei-Min Zhang and Avaroth Harindranath ,'Residual Gauge Fixing in
Light-Front QCD', hepth@xxx/9302107}
\refend   ),
so as to make the resulting commutation relation clear :
$$\L d^4 x = {1\over 2}2\big\{ \pi_A ^i d A_i - A_i d \pi_A ^i  +
\pi_\psi d\psi_+  -d \pi_\psi    \psi_+  +d\psi^{\dagger}_+
\pi_{\psi^{\dagger}} - \psi^{\dagger}_+ d \pi_{\psi^{\dagger}}
\big\} {dx^-dx_\perp  \over 2}   $$
 $$ +        {1\over 2}2\big\{\rho_A ^i d A_i- A_i d \rho_A ^i +
 \rho _\psi d \psi_- -d \rho _\psi \psi_- +d\psi^{\dagger}_-
 \rho _{\psi^{\dagger}} - \psi^{\dagger}_- d\rho_{\psi^{\dagger}}
 \big\} {dx^+dx_\perp \over 2}  $$
$$ -\H dx^+  -\K dx^- +  A_+ C_\pi d^4x + A_-  C_\rho d^4 x \eqn\cc $$
We see now that we have two types of evolutions, one along $x^+$, for
which the first term in equation  \cc\quad gives the commutation
relations along surfaces $x^+ = y^+ $ according to the form:
$$ [\xi^a , \xi^b ] = \Gamma ^{-1} _{ab}\quad a,b=1,..8\eqn\xplucc $$
with
$$\xi^1 = \pi_A ^1 , \xi^2 = \pi_A ^2 ,
\xi^3 = \pi_\psi, \xi^4 = \pi_{\psi^{\dagger}} ,
\xi^5 = A ^1, \xi^6 =A ^2,
\xi^7 = \psi_+, \xi^8 = \psi_+ ^{\dagger}\eqn\xis $$
and
$$\Gamma_{15} = \Gamma_{26} = \Gamma_{37} = \Gamma_{48}= 2 =
-\Gamma_{48} = -\Gamma_{57} =  -\Gamma_{62} = -\Gamma_{51}\eqn\gs$$
and all the other $\Gamma$'s are $0$ .
The second term in equation  \cc\quad gives the commutation
relations along surfaces $x^- = y^- $ according to the form :
$$ [\eta ^a ,\eta^b ] = \Delta ^{-1} _{ab}\quad a,b=1,..8\eqn\xmincc $$
with
$$\eta^1 = \rho_A ^1 , \eta^2 = \rho_A ^2 ,
\eta^3 = \pi_\psi, \eta^4 = \pi_{\psi^{\dagger}} ,
\eta^5 = A ^1, \eta^6 =A ^2,
\eta^7 = \psi_-, \eta^8 = \psi_- ^{\dagger}\eqn\eis $$
and
$$\Delta_{15} = \Delta_{26} = \Delta_{37} = \Delta_{48}= 2 =
-\Delta_{48} = -\Delta_{57} =  -\Delta_{62} = -\Delta_{51}\eqn\ds$$
and all the other $\Delta$'s are $0$ .
Going now to the quantum commutators, we get the following relations
for fields at equal $x^+ = y^+ $ , the usual front-form 'time' :
$$ [A^i(x^+ , x^- ,x_\perp) , \pi_A ^j (y^+, y^-, y_\perp) ] _{x^+ =
y^+} = {i \over 2}\delta(x^- - y^-)\delta^2(x_\perp - y_\perp)
\delta^{ij} \eqn\pfxplu$$
$$\big\{\psi_+(x^+,x^-,x_\perp),\pi_\psi (y^+,y^-,y_\perp)\big\}_
{x^+ = y^+} = +{i\over 2} \Lambda_+ \delta(x^- - y^-)
\delta^2(x_\perp - y_\perp) \eqn\psii$$
$$\big\{\psi_+^{\dagger}(x^+,x^-,x_\perp),
\pi_{\psi^{\dagger}} (y^+,y^-,y_\perp)\big\}_
{x^+ = y^+} = -{i\over 2} \Lambda_+ \delta(x^- - y^-)
\delta^2(x_\perp - y_\perp) \eqn\psidag $$
Thus, the  physical (quantized) degrees of freedom on $x^+ =0$ are
$A^i$ , $\psi_+$ and $\psi_+ ^\dagger$.

For fields at equal $x^- = y^-$, a new front-form 'time', we get:
$$ [A^i(x^+ , x^- ,x_\perp) , \pi_A ^j (y^+, y^-, y_\perp) ] _{x^- =
y^-} = {i \over 2}\delta(x^- - y^-)\delta^2(x_\perp - y_\perp)
\delta^{ij}\eqn\pfxmin$$
$$\big\{\psi_-(x^+,x^-,x_\perp),\rho_\psi (y^+,y^-,y_\perp)\big\}_
{x^- = y^-} = +{i\over 2} \Lambda_- \delta(x^+ - y^+)
\delta^2(x_\perp - y_\perp) \eqn\psim $$
$$\big\{\psi_-^{\dagger}(x^+,x^-,x_\perp),
\rho_{\psi^{\dagger}} (y^+,y^-,y_\perp)\big\}_
{x^- = y^-} = -{i\over 2} \Lambda_- \delta(x^+ - y^+)
\delta^2(x_\perp - y_\perp) \eqn\psimdag $$
Here, the  physical (quantized) degrees of freedom on $x^- =0$ are
$A^i$ , $\psi_-$ and $\psi_- ^\dagger$. Note that $A^+$ and $A^-$
do not enter in the list of physical (quantized) degrees of freedom.

The equations of motions are now  like in Faddeev and Jackiw [\fj]
$$ \Gamma _{ab} \partial^- \xi^b = {\partial \H \over \partial \xi^a}
\eqn\eomh $$
for the $x^+$ variation, and
$$ \Delta _{ab} \partial^- \eta^b = {\partial \K \over \partial
\eta^a}\eqn\eomk $$
for the $x^-$ variation .
For $a=5$ and $b=1$, equation \eomh\quad gives
$$ \partial_+ F^{+1}  =
+2e\psi^{\dagger}_+\gamma_0\gamma_i   \psi_-
+2e\psi^{\dagger}_-\gamma_0\gamma_i  \psi_+
  \eqn\eomf$$
which is just the equation of motion [\ema] for $\nu = 1$ .
For $a=7$ and $b=3$ we recover the equation of motion for
$\psi_+^{\dagger}$
$$ i \partial ^- \psi_+^{\dagger} = i {\partial_i \psi_-^{\dagger}
\over 2}\gamma_0\gamma_i - m\psi_-^{\dagger}{\gamma_0 \over 2}
+2 e \psi^\dagger _+ A_+  \eqn\eompsid $$
We get similar results from \eomk .

But what is the meaning of the fields $A^+$ and $A^-$ ?
They obey the following coupled set of differential equations ,according
to $C_\pi$ and $C_\rho$ :
$${1\over 2}\partial^+\partial^- A^+ - {1\over 2}(\partial^+ )^2 A^-
-(\partial^i )^2 A^+ = -\partial^i \partial^+ A^i +
 2e\psi^{\dagger}_+  \psi_+ \eqn\aplus $$
$${1\over 2}\partial^-\partial^+ A^- - {1\over 2}(\partial^- )^2 A^+
-(\partial^i )^2 A^- = -\partial^i \partial^- A^i +
 2e\psi^{\dagger}_-  \psi_- \eqn\amin $$
We've arranged the equations so that all the known fields,
the independent fields are on the right-hand side, and the 'new'
fields are on the left-hand side. The point is that these are
{\bf not} constraint equations since they are not relations between
the initial data , since neither $A^+$ nor $A^-$ get initialized
on either hyperplane! We introduce these new fields so that we
preserve Lorentz covariance and so that we have the same equations of
motion in the Euler-Lagrange case and the Hamiltonian case.

Inverting these equations, we got the following equations for $A^+$
and $A^-$ :
$$A^+ = ( (\partial^i)^2)^{-1}\partial^i\partial^+
A^i -((\partial^i)^2)^{-1}e\psi^\dagger _+\psi_+ +(\partial^+\partial^-
-(\partial^i)^2)^{-1}e \psi^\dagger _+\psi_+ $$
$$ -((\partial^i)^2)^{-1} (\partial^+\partial^-
   -(\partial^i)^2 )^{-1}(\partial^+)^2 e\psi^\dagger
_+ \psi_+ \eqn\apldef $$
$$A^- =( (\partial^i)^2)^{-1}\partial^i\partial^+
A^i -((\partial^i)^2)^{-1}e\psi^\dagger _-\psi_- +(\partial^-\partial^+
-(\partial^i)^2)^{-1}e \psi^\dagger _-\psi_- $$
$$ -((\partial^i)^2)^{-1}
(\partial^-\partial^+ -(\partial^i)^2 )^{-1}(\partial^-)^2 e\psi^\dagger
_- \psi_- \eqn\amidef $$
To fully define these fields, we need to define the operators
$((\partial^i)^2)^{-1}$ and
$(\partial^+\partial^- -(\partial^i)^2)^{-1}$ . Then we'll have
$A^+$ and $A^-$ completely determined in terms of known fields.

This is quite straight-forward. For the definition of
$(\partial^+)^{-1}$,
we use the idea of Zhang and Harindranath [\zh]
of taking anti-periodic boundary conditions for all the fields. This
determines then the definition for this operator we are considering. It
is
$${1\over \partial^ +}f(x^- ) = {1\over 2}\int {dk^+ \over 2\pi}
e^{-ik^+ x^-} \big\{
{1\over k^+ + i\epsilon} + {1\over k^+ - i\epsilon}  \big\} f(k^+ )
\eqn\dplinv $$
which leads to the following form for its square
$$({1\over \partial^ +})^2 f(x^- ) = {1\over 2}\int {dk^+ \over 2\pi}
e^{-ik^+ x^-} \big\{
{1\over k^+ + i\epsilon} + {1\over k^+ - i\epsilon}  \big\}^2 f(k^+ )
\eqn\dpllinv $$
In position space, the operator
$(\partial^+)^{-1}$,
is just the convoluted epsilon distribution [\zh], while the operator
$(\partial^+)^{-2}$,
becomes
$${1\over 2}\int_{-\lambda} ^{\lambda} dx^- \epsilon(x^- - x'^-)
\epsilon(x^- - x"^- ) = -|x'^- - x"^-| + \lambda \eqn\conv $$
As Zhang and Harindranath point out, it is crucial that we take
this definition in getting a consistent specification of the front-form
singularity $k^+ = 0$.

We treat the sibling operator
$(\partial^-)^{-1}$,
like we did $(\partial^+)^{-1}$. We have
$${1\over \partial^ -}f(x^+ ) = {1\over 2}\int {dk^- \over 2\pi}
e^{-ik^- x^+} \big\{
{1\over k^- + i\epsilon} + {1\over k^- - i\epsilon}  \big\} f(k^- )
\eqn\dmiinv $$
This  leads to the following form for its square
$$({1\over \partial^ -})^2 f(x^+ ) = {1\over 2}\int {dk^- \over 2\pi}
e^{-ik^- x^+} \big\{
{1\over k^- + i\epsilon} + {1\over k^- - i\epsilon}  \big\}^2 f(k^- )
\eqn\dmiiinv $$
Just like above, the operator
$(\partial^-)^{-1}$,
is just the convoluted epsilon distribution [\zh], while the operator
$(\partial^-)^{-2}$,
becomes
$${1\over 2}\int_{-\lambda}^{\lambda} dx^- \epsilon(x^+ - x'^+)
\epsilon(x^+ - x"^+ ) = -|x'^+ - x"^+| + \lambda \eqn\convv $$

The other two operators are simpler to define.
$((\partial^i)^2)^{-1}$
 becomes
$$((\partial^i)^2)^{-1} f(x_\perp) =\int {d^2 k_\perp \over (2\pi)^2}
{e^{+i k_\perp x_\perp} \over -k_\perp ^2 + i\epsilon} f(k_\perp)
\eqn\dperpinv $$
while
$(\partial^+ \partial^- -(\partial^i)^2)^{-1}$
becomes
$$(\partial^+ \partial^- -(\partial^i)^2)^{-1}=\int {d^2 k_\perp
\over (2\pi)^2 } {dk^+ dk^- \over (2\pi)^2} {e^{-i k x} \over
4k^+ k^- - k_\perp ^2 +i\epsilon} f(k^+, k^-, k_\perp)
\eqn\fullinv $$ .

\chapter{Quantization of the Fields}
Now that we have the commutation relations, we are ready to define
the fields $A^i$ and $\psi$.  According to [\ivpc ], using two null
hyperplanes, the initial data must be specified on each of the
hyperplanes as well as on their intersection .
In this case,    we will have  initialization
 on the two surfaces $x^+ =0$ and $x^- =0$ . We will
require, though, that on the intersection of these surfaces, at
$x^+ =x^- = 0$ these fields satisfy certain consistency conditions.
This works out as follows.

On $x^+ =0$ we have then :
$$ A_i(x^+ = 0, x^- , x_\perp ) = \int {d^2 k_\perp \over (2\pi)^3 }
{dk^+ \over 2 k^+ } \big\{ \epsilon_i (k^+ , k_\perp)
 a(k^+ , k_\perp ) e^{-i  k . x} + \epsilon_i ^* (k^+ ,k_\perp)
 a^{\dagger}(k^+ , k_\perp ) e^{+i  k . x} \big\} \eqn\aplu  $$
$$\psi_+(x^+ = 0, x^- , x_\perp) = \int {d^2 k_\perp \over (2\pi)^3 }
{dk^+ \over 2 k^+ }\sum_{\lambda} \big\{ b(k^+ , k_\perp ) u_+(k^+ ,
k_\perp, \lambda) e^{-i k . x} $$
$$ + d^{\dagger}(k^+ , k_\perp)
v_+(k^+ , k_\perp, \lambda) e^{+i k . x} \big\} \eqn\psipl $$
In this case, $i k . x = i k^+ x^- - i k_\perp . x_\perp $   and
the polarization vector is $\epsilon_i (k^+ ,k_\perp) $.

On the other hyperplane, $x^- = 0$ we get similar forms:
$$ A_i(x^- = 0, x^+ , x_\perp ) = \int {d^2 k_\perp \over (2\pi)^3 }
{dk^- \over 2 k^- } \big\{ \hat \epsilon_i (k^- , k_\perp)
\hat a(k^- , k_\perp ) e^{-i \hat k . x} +\hat\epsilon_i ^*(k^- ,k_\perp)
\hat a^{\dagger}(k^- , k_\perp ) e^{+i \hat k . x} \big\} \eqn\phimi $$
$$\psi_-(x^- = 0, x^+ , x_\perp) = \int {d^2 k_\perp \over (2\pi)^3 }
{dk^- \over 2 k^- }\sum_{\mu} \big\{ \hat b(k^- , k_\perp ) u_-(k^- ,
k_\perp, \mu) e^{-i \hat k . x} $$
$$ + \hat d^{\dagger}(k^- , k_\perp)
v_-(k^- , k_\perp, \mu) e^{+i \hat k . x} \big\} \eqn\psimi $$
Here  , $i \hat k . x = i k^- x^+ - i k_\perp . x_\perp $ .

We require now that the fields  be consistent at $x^+ = x^- = 0$. This
means  that we have
$$ A_i (x^+ = 0, x^- = 0, x_\perp) = A_i (x^- = 0, x^+ = 0, x_\perp)
\eqn\aiai $$
 This implies
$$  \int {d^2 k_\perp \over (2\pi)^3 }
{dk^+ \over 2 k^+ } \big\{ \epsilon_i (k^+ ,k_\perp)
a(k^+ , k_\perp ) e^{+i k_\perp . x_\perp} +\epsilon_i (k^+ ,k_\perp)^*
a^{\dagger}(k^+ , k_\perp ) e^{-i k_\perp . x_\perp} \big\} $$
$$  = \int {d^2 k_\perp \over (2\pi)^3 }{dk^- \over 2 k^- }
\big\{ \hat\epsilon_i (k^- ,k_\perp)
\hat a(k^- , k_\perp ) e^{+i k_\perp . x_\perp} +
\hat\epsilon_i (k^- ,k_\perp)^* \hat a^{\dagger}
(k^- , k_\perp ) e^{-i k_\perp . x_\perp}\big\}\eqn\aadag $$
As $k^+$ and $k^-$ are just dummy variables here, we get that
$$a(k^+ , k_\perp) = \hat a(k^+  , k_\perp),\quad a^{\dagger}
(k^+ , k_\perp) = \hat a^{\dagger}(k^+  , k_\perp) \eqn\aad$$
as well as
$$\epsilon_i (k^+ ,k_\perp) = \hat\epsilon_i (k^+ ,k_\perp)\eqn\eps$$
and we need to point out that the  variables are the {\bf same} for
both creation operators.
So this means that
$$a(k^+ , k_\perp) \ne \hat a(k^- , k_\perp),\quad
\epsilon_i (k^+ , k_\perp) \ne \hat\epsilon_i (k^- , k_\perp) \eqn\aane$$
hence the field  $A_i $ has different effects on the two surfaces.
On $x^+ =0$, $A_i(x^+ = 0, x^- , x_\perp)$
creates or destroys vector quanta with momentum
$ k = (k^+ , k_\perp) $  and polarization $\epsilon_i (k^+ ,k_\perp)$.
On $x^- =0$, $A_i(x^- =0, x^+ , x_\perp)$
creates or destroys quanta with momentum $\hat k = (k^- , k_\perp)$
and polarization $\epsilon_i (k^- ,k_\perp)$ .
\REF\fourier{Note that the Fourier decomposition goes through eventhough
we have two 'times.' The point is that on each constant surface, be it
$x^+ =0$ or $x^- =0$, there is only {\bf one} front-form time,
so that there
is a consistent  definition of a Fourier decomposition. On each of
these surfaces,  space-time looks like $2+1$ Minkowski space-time.}
\refend

The analysis for the fermion fields goes through
just like in the previous paper [\pffq] .

What about the fields $A^+$ and $A^-$ ? As mentioned in the previous
section, as these fields are not initialized on any of the surfaces,
they do not constitute constraints. We have solved equations [\aplus]
and [\amin] in terms of the independent degrees of freedom $A^i$ and
$\psi_+$ , $\psi_-$  in equations [\apldef] and [\amidef].
We get the following :
$$A^+ (x^+  , x^- , x_\perp)
= ( (\partial^i)^2)^{-1}\partial^i\partial^+
A^i (x^+ =0, x^- ,x_\perp) $$
$$-((\partial^i)^2)^{-1}e\psi^\dagger _+\psi_+ (x^+ =0, x^- ,x_\perp)
+(\partial^+\partial^- -(\partial^i)^2)^{-1}e \psi^\dagger _+
\psi_+ (x^+ =0, x^- ,x_\perp) $$
$$ -((\partial^i)^2)^{-1} (\partial^+\partial^-
   -(\partial^i)^2 )^{-1}(\partial^+)^2 e\psi^\dagger
_+ \psi_+ (x^+ = 0, x^- , x_\perp) \eqn\apldf $$
$$A^- (x^- , x^+ ,x_\perp)
= ( (\partial^i)^2)^{-1}\partial^i\partial^+
A^i (x^- =0, x^+ ,x_\perp) $$
$$-(\partial^i)^2)^{-1}e\psi^\dagger _-\psi_- (x^- =0, x^+ ,x_\perp)
+(\partial^-\partial^+ -(\partial^i)^2)^{-1}e \psi^\dagger _-
\psi_- (x^- =0, x^+ ,x_\perp) $$
$$ -((\partial^i)^2)^{-1}
(\partial^-\partial^+ -(\partial^i)^2 )^{-1}(\partial^-)^2 e\psi^\dagger
_- \psi_- (x^- =0,x^+ ,x_\perp)  \eqn\amidf $$
where we use the definitions [\dperpinv], [\fullinv], [\dpllinv] and
[\dmiiinv] .

So all the fields coming in the definition of $QED$ are defined and
$A^+$ or $A^-$ do not represent new modes or new quanta.
It is important to point out here that our gauge field $A$ has only
{\bf two} physical degrees of freedom , $A^i ,  i=1,2$.
The fields $A^+$ and $A^-$ are needed to guarantee Lorentz covariance,
but are not gotten from some constraint equations.

Let us point out that these equations are different in nature than
similar  equations one gets in the case of constraint quantization.
In the constrained case, one needs to
solve the constraint equation {\bf before} quantization. This is often
hard and sometimes impossible analytically. Here, we have already
quantized our theory and are computing new fields, so we are past the
quantization stage. The quantization procedure seems easier
in this approach than in the constrained approaches [\cd], [\gt],
[\gov], [\ht] .
\chapter{Parity in Front-Form Quantization}
We are ready now to study how the fields $A^i$  ,  $\psi_+$ and
$\psi_-$ transform under parity. For this we use
(Bjorken and Drell for instance
\REF\bd{James  D. Bjorken and Sidney D. Drell, 'Relativistic Quantum
Fields,' Chapter 15 , McGraw-Hill, San Francisco, 1965}
\refend  ) :
$$ \P A^i(x^+ , x^- , x_\perp) \P^{-1} = - A^i(x^- , x^+ , -x_\perp)
\eqn\scpar $$
since under parity $(x^+ , x^- , x_\perp) \rightarrow (x^- , x^+ ,
-x_\perp)$  and the vector field has negative intrinsic parity.
For the vector field we get
$$ \P A^i(x^+ = 0, x^- , x_\perp )\P^{-1} =\P
\int {d^2 k_\perp \over (2\pi)^3 }{dk^+ \over 2 k^+ } \big\{
\epsilon_i (k^+ ,k_\perp)a(k^+ , k_\perp ) e^{-i k . x} $$
$$ +\epsilon_i ^* (k^+ ,k_\perp)
a^{\dagger}(k^+ , k_\perp ) e^{+i k . x} \big\}\P^{-1} \eqn\aipa $$
This becomes
$$ \P A^i(x^+ = 0, x^- , x_\perp )\P^{-1} =
   \int {d^2 (-k_\perp) \over (2\pi)^3 }
{dk^- \over 2 k^- } \big\{-\epsilon_i (k^- ,-k_\perp)
a(k^- ,-k_\perp ) e^{-i k'. x'} $$
$$ -\epsilon_i ^* (k^- ,-k_\perp)
a^{\dagger}(k^- ,-k_\perp ) e^{+i k'. x'} \big\}\eqn\aipb $$
if
$$\P a(k^+ ,k_\perp) \P^{-1} = a(k^- , -k_\perp),\quad \P a^{\dagger}
(k^+ ,k_\perp) \P^{-1} = a^{\dagger}(k^- , -k_\perp) $$
and
$$ \P\epsilon_i (k^+ ,k_\perp)\P^{-1} = -\epsilon_i (k^-, -k_\perp)
\eqn\apa $$
and $i k' . x' = ik^- x^+ - i k_\perp x_\perp$ . Redefining variables
$(k^-, -k_\perp)\rightarrow (l^-,l_\perp)$, we get the result
$$ \P  A^i(x^+ =0, x^- , x_\perp) \P^{-1} =-A^i(x^- =0, x^+ , -x_\perp)
  \eqn\aipc$$
Let us consider the fermion fields now. In this case we have the same
result of the previous paper [\pffq]
$$ \P \psi (x^+ , x^- , x_\perp) \P^{-1} = \gamma_0 \psi (x^- , x^+ ,
-x_\perp) \eqn\psipa $$
and   we expect that fields defined on $x^+$ will be mapped into
fields defined on $x^-$ by parity. Indeed, that is what we find for
$\psi_+$ .

We derive now these relations for arbitrary $x^+$ and $x^-$.
Note that for the $x^+$  evolution  we have
$$ A^i(x^+ , x^- , x_\perp)  = e^{-iP^- x^+} A^i
(x^+ =0 , x^- , -x_\perp) \eqn\evolai $$
or
$$\psi_- (x^+ , x^- , x_\perp)  = e^{-iP^- x^+} \psi_-
(x^+ =0 , x^- , -x_\perp) \eqn\evolpsi $$
so that the parity-transformed field is
$$ \P A^i(x^+ , x^- , x_\perp)\P^{-1}  = \P e^{-iP^- x^+} \P^{-1}
\P A^i(x^+ =0 , x^- , -x_\perp)\P^{-1} \eqn\evolaib$$
which becomes
$$ \P A^i(x^+ , x^- , x_\perp)\P^{-1}  = e^{-iP^+ x^-} A^i
(x^- =0 , x^+ , -x_\perp) \eqn\evolaic$$
since
$$\P P^{-} \P^{-1} = \P \int \H \P^{-1} = \int \K = P^{+} \eqn\parh$$
by use of the equations \pmin\quad and \pplu.
A similar result holds for the fermion case.

We also get the generator  of $x^-$  evolutions to transform properly
as well since
$$\P P^{+} \P^{-1} = \P \int \K \P^{-1} = \int \H = P^{-} \eqn\park$$
again, by use of equations \pplu\quad and \pmin.

Since now the generators of evolution along $x^+$ and $x^-$ ($\H$ and
$\K$ respectively), transform properly under parity ,
we can evolve the
parity relations obtained at $x^+ =0$ and $x^- =0$ to
relations for arbitrary $x^+$ and $x^-$. For the vector case we get
$$ \P A^i(x^+ , x^- , x_\perp) \P^{-1} =-A^i(x^- , x^+ , -x_\perp)
  \eqn\aipd$$
as expected from previous work [\pclcq] .

For the fermion  case, we get [\dm]
$$ \P \psi_+(x^+ , x^- , x_\perp) \P^{-1} = \gamma_0 \psi_-
(x^- , x^+ , -x_\perp) \eqn\psiph $$
which show very clearly that parity maps independent fields on
$x^+ =0$ [$\psi_+(x^+ =0, x^- , x_\perp)$] ,
to independent fields on $x^- =0$  [$\psi_-(x^- =0, x^+ , x_\perp) $] ,
demonstrating the it is crucial that
we take {\bf both} $x^+ =0$ and $x^- =0$ as quantizing surfaces if we
desire to have fields with parity as an explicit symmetry as already
noted [\pffq].

Thus far we have looked at transformation properties of independent
fields on $x^+ =0$ . It is quite straightforward to show that we get
similar results for the fields which are initialized on $x^- =0$ :
$$ \P A^i(x^- , x^+ , x_\perp) \P^{-1} =-A^i(x^+ , x^- , -x_\perp)
  \eqn\aipe$$
for the vector field  and
$$ \P \psi_-(x^- , x^+ , x_\perp) \P^{-1} = \gamma_0 \psi_+
(x^+ , x^- , -x_\perp) \eqn\psipg $$
for the fermion field [\dm] .

Let us examine the parity transformation properties of the fields
$A^+$ and $A^-$. It is a straightforward exercise to  check, using
equations [\apldf] and [\amidf] that we get
$$\P A^+ (x^+=0, x^- ,x_\perp)\P^{-1} =  A^- (x^- = 0, x^+ , -x_\perp)
\eqn\papl $$
due to the transformation properties of the fields $A^i$
and $\psi_+$.   We likewise get
$$\P A^+ (x^+, x^- ,x_\perp)\P^{-1} =  A^- (x^- , x^+ , -x_\perp)
\eqn\paplb$$
for arbitrary $x^+$.

For the other field $A^-$, results come out as expected as well
$$\P A^- (x^-=0, x^+ ,x_\perp)\P^{-1} =  A^+ (x^+ = 0, x^- , -x_\perp)
\eqn\pami $$
due to the transformation properties of the fields $A^i$
and $\psi_-$.   We likewise get
$$\P A^- (x^-, x^+ ,x_\perp)\P^{-1} =  A^+ (x^+ , x^- , -x_\perp)
\eqn\pamib$$
for arbitrary $x^-$.
This completes our demonstration that fields
defined on $x^+ =0$ and $x^- =0$ transform properly under parity, and
    define $QED$  consistently.

\chapter{Acknowledgements}

I  would like to thank Dharam Ahluwalia and Prof. Stanley Brodsky for
discussions and to thank Prof. Richard Blankenbecler
for his continuing support.
\endpage
\refout
\end